\documentclass[twocolumn,showpacs,prb,amsmath,amssymb]{revtex4}

\usepackage{graphicx}% Include figure files
\usepackage{dcolumn}% Align table columns on decimal point
\usepackage{bm}% bold math

\begin{document}

\preprint{APS/123-QED}

\title{Orbital Degeneracy and Peierls Instability in Triangular Lattice Superconductor Ir$_{1-x}$Pt$_x$Te$_2$}% Force line breaks with \\

\author{Daiki Ootsuki$^1$}
\author{Yuki Wakisaka$^2$}
\author{Sunseng Pyon$^3$}
\author{Kazutaka Kudo$^3$}
\author{Minoru Nohara$^3$}
\author{Masashi Arita$^4$}
\author{Hiroaki Anzai$^4$}
\author{Hirofumi Namatame$^4$}
\author{Masaki Taniguchi$^{4,5}$}
\author{Naurang L. Saini$^{6,2}$}
\author{Takashi Mizokawa$^{2,1}$}

\affiliation{$^1$Department of Physics, University of Tokyo,5-1-5 Kashiwanoha, Chiba 277-8561, Japan}
\affiliation{$^2$Department of Complexity Science and Engineering, University of Tokyo,5-1-5 Kashiwanoha, Chiba 277-8561, Japan}
\affiliation{$^3$Department of Physics, Okayama University, Kita-ku, Okayama 700-8530, Japan}
\affiliation{$^4$Hiroshima Synchrotron Radiation Center, Hiroshima University, Higashi-hiroshima 739-0046, Japan}
\affiliation{$^5$Graduate School of Science, Hiroshima University, Higashi-hiroshima 739-8526, Japan}
\affiliation{$^6$Department of Physics, University of Roma "La Sapienza" Piazalle Aldo Moro 2, 00185 Roma, Italy}

\date{\today}% It is always \today, today,

\begin{abstract}
We have studied electronic structure of triangular lattice Ir$_{1-x}$Pt$_x$Te$_2$ superconductor 
using photoemission spectroscopy and model calculations. Ir $4f$ core-level photoemission spectra 
show that Ir $5d$ $t_{2g}$ charge modulation established in the low temperature phase of IrTe$_2$ 
is suppressed by Pt doping. This observation indicates that the suppression of charge modulation 
is related to the emergence of superconductivity. Valence-band photoemission spectra of IrTe$_2$ 
suggest that the Ir $5d$ charge modulation is accompanied by Ir $5d$ orbital reconstruction. 
Based on the photoemission results and model calculations, we argue that the orbitally-induced 
Peierls effect governs the charge and orbital instability in the Ir$_{1-x}$Pt$_x$Te$_2$.
\end{abstract}

\maketitle

\section{Introduction}

Fe pnictides and chalcogenides, including LaFeAsO$_{1-x}$F$_x$ \cite{1,2} and 
FeSe$_{1-x}$Te$_x$,\cite{3,4} show an interesting interplay between superconductivity 
and magnetism which is deeply related to the multi-band structure derived 
from the Fe $3d$ orbitals. Recently, Pyon {\it et al.} have reported superconductivity 
in triangular lattice IrTe$_2$ [see Fig. 1(a)] when Pt is substituted for Ir.\cite{5}
Interestingly, the parent material IrTe$_2$ shows a structural phase transition 
at $\sim$ 250 K which is probably due to Ir $5d$ $t_{2g}$ orbital order or bond order.\cite{6}
The Pt doping suppresses the static orbital or bond order, and the superconductivity appears around 
the quantum critical point where the orbital or bond order disappears. The comparison of 
superconductivity in triangular lattice IrTe$_2$ and square lattice Fe pnictides$/$chalcogenides 
is very interesting and may provide clues to understand the mechanism of superconductivities 
in these materials.
In addition, orbital effect on Fe pnictides$/$chalcogenides superconductors is currently 
under hot debate. Particularly, the Fe $3d$ $yz/zx$ orbital degeneracy in the tetragonal phase
of Fe pnictides$/$chalcogenides is similar to the Ir $5d$ $yz/zx$ orbital degeneracy
of the trigonal phase of Ir$_{1-x}$Pt$_x$Te$_2$ although the Ir $5d$ spin-orbit interaction
provides some differences. Therefore, study of orbital effect on Ir$_{1-x}$Pt$_x$Te$_2$ 
is highly interesting and important.

IrTe$_2$ and PtTe$_2$, which have $d^5$ configuration of Ir$^{4+}$ ions and $d^6$ configuration 
of Pt$^{4+}$ ions respectively, crystallize into the CdI$_2$-type structure as shown in Fig. 1(a). 
IrTe$_2$ exhibits a structural phase transition at $\sim$ 250 K from the trigonal (P3m-1) 
to a monoclinic (C2/m) structure, accompanied by temperature dependent anomalies of electrical resistivity 
and magnetic susceptibility. On the other hand, PtTe$_2$ with the trigonal structure does not 
exhibit the structural phase transition.\cite{6} While no superconductivity has been reported 
for IrTe$_2$ and PtTe$_2$,\cite{8,9} trigonal Ir$_{1-x}$Pt$_x$Te$_2$ shows the superconductivity 
in the vicinity of the monoclinic phase.\cite{5} The structural phase transition of IrTe$_2$ 
would be related to the orbital degeneracy of the Ir$^{4+}$($d^5$) state as shown in Fig. 1(b) 
and has similarity to the structural transition of spinel-type CuIr$_2$S$_4$.\cite{10,11,12}
In this paper, we report core-level and valence-band photoemission spectroscopy of triangular 
lattice Ir$_{1-x}$Pt$_x$Te$_2$ superconductor. Photoemission results and model calculations 
indicate that the orbitally-induced Peierls effect plays an important role in the charge-orbital 
instability and the superconductivity of Ir$_{1-x}$Pt$_x$Te$_2$.

\begin{figure}
\includegraphics[width=9cm]{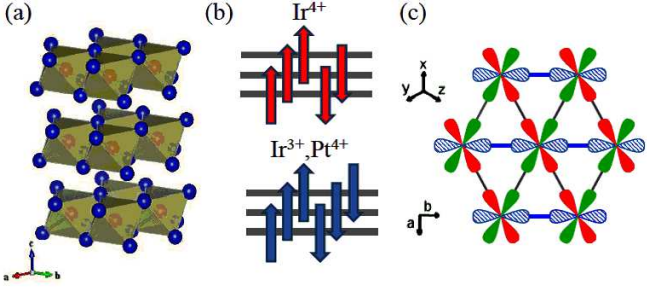}
\caption{(color online) (a) Crystal structure of IrTe$_2$ that was visualized by 
using the software package VESTA. \cite{7} The IrTe$_6$ octahedra share their edges 
and form the IrTe$_2$ triangular lattice layer. (b) Electronic configurations of 
Ir$^{4+}$($d^5$) and Ir$^{3+}$, Pt$^{4+}$($d^6$). (c) Ir $5d$ t$_{2g}$ orbitals 
on the triangular lattice. The thicker lines indicate shorter Ir-Ir bonds which are 
due to bond formation of the shaded Ir $5d$ yz orbitals.}
\end{figure}

\section{Methods}

\subsection{Experiment}

The polycrystalline samples of Ir$_{1-x}$Pt$_x$Te$_2$ ($x=0.00$, $0.03$, and $0.04$) 
were prepared as reported in ref. 5.\cite{5} IrTe$_2$ is nonsuperconducting while 
Ir$_{1-x}$Pt$_x$Te$_2$ ($x$= 0.03, and 0.04) are superconducting with $T_c$ = 3.1 K and 2.9 K 
respectively. The photoemission measurements were performed at beamline 9A, 
Hiroshima Synchrotron Radiation Center using a SCIENTA R4000 analyzer with 
circularly polarized light. The total energy resolution was set to 8 meV for 
the selected excitation energy of $h\nu=10$ eV. The angular resolution was set to 
$\sim$ 0.3$^{\circ}$ that gives the momentum resolution of $\sim$ 0.01 A$^{-1}$ for $h\nu=10$ eV. 
The circular polarization of the incident beam is 50$^o$ off the sample surface. 
The base pressure of the spectrometer was in the $10^{-9}$ Pa range. 
The polycrystalline samples of Ir$_{1-x}$Pt$_x$Te$_2$ were fractured at $300$ K 
under the ultrahigh vacuum and the spectra were acquired within 6 hours 
after the fracturing. The x-ray photoemission spectroscopy (XPS) was carried out
at 300 K and 40 K using JEOL JPS9200 analyzer. Monochromatic Al K$\alpha$ ($1486.6$ eV) was 
used as x-ray source. The total energy resolution was about $0.6$ eV. 
The base pressure of the chamber was in the $10^{-7}$ Pa range. The binding energy 
was calibrated using the Au $4f$ core level of the gold reference sample. 
We fractured the polycrystalline samples of Ir$_{1-x}$Pt$_x$Te$_2$ at $300$ K 
for the XPS measurements.

\subsection{Calculation}

The electronic structure of IrTe$_2$ was analyzed using a tight-binding model
with Ir 5$d$ and Te 5$p$ orbitals.
The tight-binding Hamiltonian of the multi-band model \cite{13} is given by 
\begin{eqnarray*}
H = H_p + H_d + H_{pd},
\end{eqnarray*}
\begin{eqnarray*}
H_p = \sum_{k,l,\sigma}
\epsilon^p p^+_{k,l\sigma}p_{k,l\sigma}
+ \sum_{k,l>l',\sigma}
V^{pp}_{k,ll'} p^+_{k,l\sigma}p_{k,l'\sigma} + H.c.,
\end{eqnarray*}
\begin{eqnarray*}
H_d & = &\sum_{i,m\sigma}\epsilon_d d^+_{i,m\sigma}d_{i,m\sigma}
+\sum_{i,m,m',\sigma,\sigma'}
h^{d}_{mm'\sigma\sigma'} d^+_{i,m\sigma}d_{i,m'\sigma'}
\end{eqnarray*}
\begin{eqnarray*}
+ \sum_{k,m>m',\sigma}
V^{dd}_{k,mm'} d^+_{k,m\sigma}d_{k,m'\sigma} + H.c.,
\end{eqnarray*}
\begin{eqnarray*}
H_{pd} = \sum_{k,m,l,\sigma} V^{pd}_{k,lm}
d^+_{k,m\sigma}p_{k,l\sigma} + H.c.
\end{eqnarray*}

\noindent Here, $d^+_{i,m\sigma}$ are creation operators for the Ir 5$d$ 
electrons with orbital $m$ and spin $\sigma$ at site $i$. 
$d^+_{k,m\sigma}$ and $p^+_{k,l\sigma}$ are creation 
operators for Bloch electrons with momentum $k$ which are 
constructed from the $m$-th component of the Ir 5$d$ orbitals 
and from the $l$-th component of the Te 5$p$ orbitals, respectively.
$h^{d}_{mm'\sigma\sigma'}$ represents the ligand field splitting and 
the atomic spin-orbit interaction for the Ir 5$d$ orbitals.
The transfer integrals $V^{pd}_{k,lm}$  between the Ir 5$d$ and Te 5$p$ orbitals 
are given by Slater-Koster parameters (pd$\sigma$) and (pd$\pi$) which are set to -2.0 eV and 
0.9 eV for the undistorted trigonal structure. Also the Te 5$p$-Te 5$p$ transfer integrals
$V^{pp}_{k,ll'}$ are given by (pp$\sigma$) and (pp$\pi$) of 0.6 eV and -0.15 eV, 
and the Ir 5$d$-Ir 5$d$ transfer integrals $V^{dd}_{k,mm'}$ are given by
(dd$\sigma$) and (dd$\pi$) of -0.4 eV and 0.15 eV for the undistorted trigonal structure. 
The magnitude of the Ir 5$d$ spin-orbit interaction is set to 0.6 eV. 
When the trigonal structure is distorted, the Slater-Koster parameters are modified 
using the Harrison's rule. \cite{13} 
The effect of orbital or bond order can be examined by introducing the $5$\% 
bond compression along the b-axis which is consistent with the experimental value.
The transfer integrals along the b-axis are enhanced by the bond compression
and, consequently, the degeneracy of the Ir 5$d$ $t_{2g}$ bands can be removed.
This can be viewed as a kind of band Jahn-Teller effect. The band Jahn-Teller
effect modifies the Fermi surface geometry to cause Peierls instability 
in the orbitally-induced Peierls mechanism.
The Te 5$p$-to-Ir 5$d$ charge-transfer energy $\Delta$ ($=\epsilon^d-\epsilon^p$) 
is taken as an adjustable parameter to reproduce the spectral weight suppression 
at -0.1 eV by the lattice distortion.

\section{Results and Discussion}

Ir $4f$ and Te $3d$ core-level photoemission spectra of Ir$_{1-x}$Pt$_x$Te$_2$ 
($x=0.00$, $0.03$,and $0.04$) are displayed in Fig. 2. For $x$ = 0.03 and 0.04, 
the Te $3d$ peaks are accompanied by shoulders at $\sim$ 575 eV which can be 
attributed to Te impurities at grain boundaries. The absence of this shoulder 
for IrTe$_2$ indicates that the photoemission results for IrTe$_2$ are highly reliable. 
As for the Pt doped samples, since the main Te $3d$ peaks representing the bulk 
Ir$_{1-x}$Pt$_x$Te$_2$ are still dominant, the Ir 4$f$ and valence-band 
photoemission results can be used to discuss the bulk electronic structure.  
As shown in Fig. 2(a), the Ir $4f$ peak width of IrTe$_2$ slightly increases 
in going from $300$ K to $40$ K while those of the Pt doped samples as well as 
the Te $3d$ peaks do not show any such changes with temperature.
The increase of peak width indicates that the density of Ir $5d$ $t_{2g}$ electrons 
is modulated in the low temperature phase of IrTe$_2$. Here, it should be noted that
the Ir $4f$ peak width increase of IrTe$_2$ is comparable to that of CuIr$_2$S$_4$
in which the octamer Ir$^{3+}$/Ir$^{4+}$ charge ordering was established \cite{11} 
and the charge difference between the Ir$^{3+}$ site and the Ir$^{4+}$ site 
was observed in the Ir $4f$ XPS.\cite{12}

On the other hand, the Ir $4f$ peak width of the Pt doped samples does not
change appreciably with temperature, indicating that the Ir $5d$ charge modulation 
is suppressed by the Pt doping. Also the Ir $4f$ peaks of the Pt doped samples 
have asymmetric line shape due to the increase of conduction electron by the Pt doping.
Interestingly, the Ir $4f$ binding energy of IrTe$_2$ is smaller than that of CuIr$_2$S$_4$ 
as shown in Fig. 2(a), suggesting that the actual number of Ir 5$d$ electrons of  IrTe$_2$ 
(formally Ir$^{4+}$) is larger than that of CuIr$_2$S$_4$ (formally Ir$^{3.5+}$). 

\begin{figure}
\includegraphics[width=9cm]{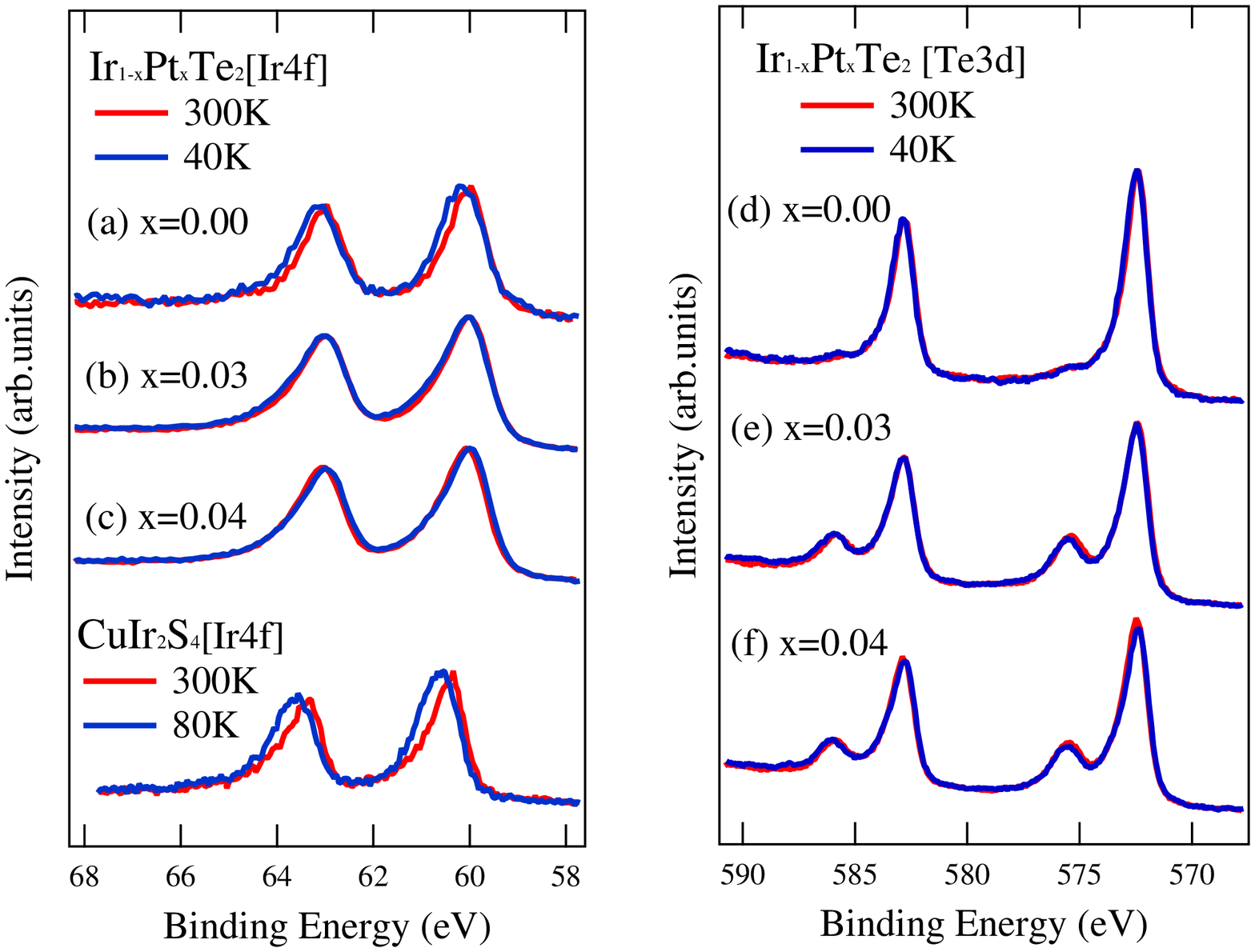}
\caption{(color online) Ir $4f$ core-level photoemission spectra of Ir$_{1-x}$Pt$_x$Te$_2$ 
for (a) $x=0.00$, (b) $x=0.03$, and (c) $x=0.04$ compared to that of CuIr$_2$S$_4$.\cite{12}
Te $3d$ core-level photoemission spectra for (d) $x=0.00$, (e) $x=0.03$, and (f) $x=0.04$.}
\end{figure}

Valence-band photoemission spectra of Ir$_{1-x}$Pt$_x$Te$_2$ ($x=0.00$, $0.03$, and $0.04$) are displayed 
in Figs. 3(a)-(c). In IrTe$_2$, across the orbital or bond order temperature at $\sim$ 250 K, 
the spectral weight around -0.1 eV is suppressed instead of that at the Fermi level [Fig.3 (a)]. 
The spectral weight suppression seems to rapidly disappear with the Pt doping. In order to 
clarify spectral weight change, we divided the photoemission spectra of Ir$_{1-x}$Pt$_x$Te$_2$ 
by the Fermi-Dirac function convoluted with a Gauss function of FWHM of $8$ meV as shown in 
Figs. 3(d)-(f). The spectral weight around -0.1 eV of IrTe$_2$ is suppressed in the low temperature phase. 
On the other hand, the spectral weight at the Fermi level is almost preserved across the structural 
transition at $\sim$ 250 K, consistent with the good metallic behavior of the orbital 
or bond order state.\cite{5,6} However, this result apparently contradicts with the dramatic suppression 
of magnetic susceptibility below $\sim$ 250 K.\cite{5}

\begin{figure}
\includegraphics[width=9cm]{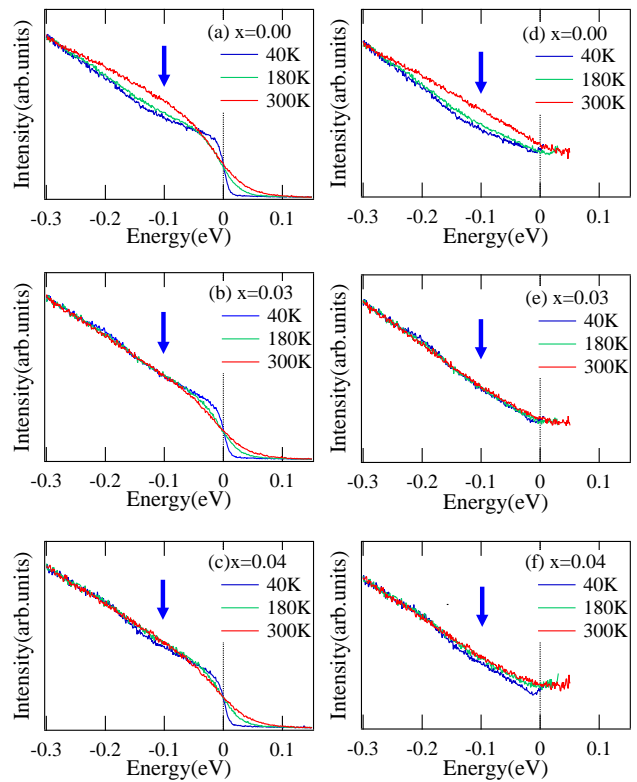}
\caption{(color online) Temperature dependent photoemission spectra near the Fermi level of
Ir$_{1-x}$Pt$_x$Te$_2$ for (a) $x=0.00$, (b) $x=0.03$, and (c) $x=0.04$. The spectra are taken 
at photon energy of $10$ eV. Temperature dependence of the photoemission spectra divided by the broadened
Fermi-Dirac function of Ir$_{1-x}$Pt$_x$Te$_2$ for (d) $x=0.00$, (e) $x=0.03$, and (f) $x=0.04$.}
\end{figure}

The spectral weight suppression around -0.1 eV of IrTe$_2$ would be consistent with band narrowing 
due to a kind of band Jahn-Teller effect caused by the bond compression. Under the bond compression 
along the b-axis (namely, the orbital or bond order along the b-axis), the Ir $5d$ yz band width 
along the b-axis is increased and the Ir $5d$ xy and zx band width 
is decreased [see Fig. 1(c)]. If the narrow Ir $5d$ xy and zx bands become fully occupied and 
their tops are located below $\sim$ -0.1 eV from the Fermi level, the density of states 
down to $\sim$ -0.1 eV is expected to be suppressed. This situation can be demonstrated 
by the tight binding calculation for a multi-band model including the Ir $5d$ and Te $5p$ orbitals 
with realistic transfer integrals and spin-orbit interactions.
Using $\Delta$ of 1.0 eV, the Ir $5d$ holes are accommodated in the Ir $5d$ yz orbitals 
and the density of states from the Fermi level to -0.1 eV is actually suppressed 
as shown in Fig. 4(a). Figure 4(a) also shows the calculated results without
the spin-orbit interaction. Although the effect of the spin-orbit interaction 
is rather small near the Fermi level (the geometry of the Fermi surface is
also the same with and without the spin-orbit interaction), the density of
states below $\sim$ -0.1 eV is affected by the spin-orbit interaction.
The calculated results for $\Delta$ of 2.0 eV and -2.0 eV are displayed 
in Fig. 4(b). The calculated spectral change is too large for $\Delta$ 
larger than 2.0 eV, and it is too small for $\Delta$ 
smaller than -2.0 eV. The $\Delta$ value close to 0 eV indicates that 
the Ir 5$d$-Te 5$p$ hybridization is substantial although
the charge density wave formation of the low temperature phase manifests 
only in the Ir 4$f$ core level (not in the Te 3$d$ core level).
Here, it should be noted that the orbitally-ordered (or bond-ordered) state 
is unstable without the lattice distortion in the present model calculation
probably because the electron-electron interaction is not included.

\begin{figure}
\includegraphics[width=9cm]{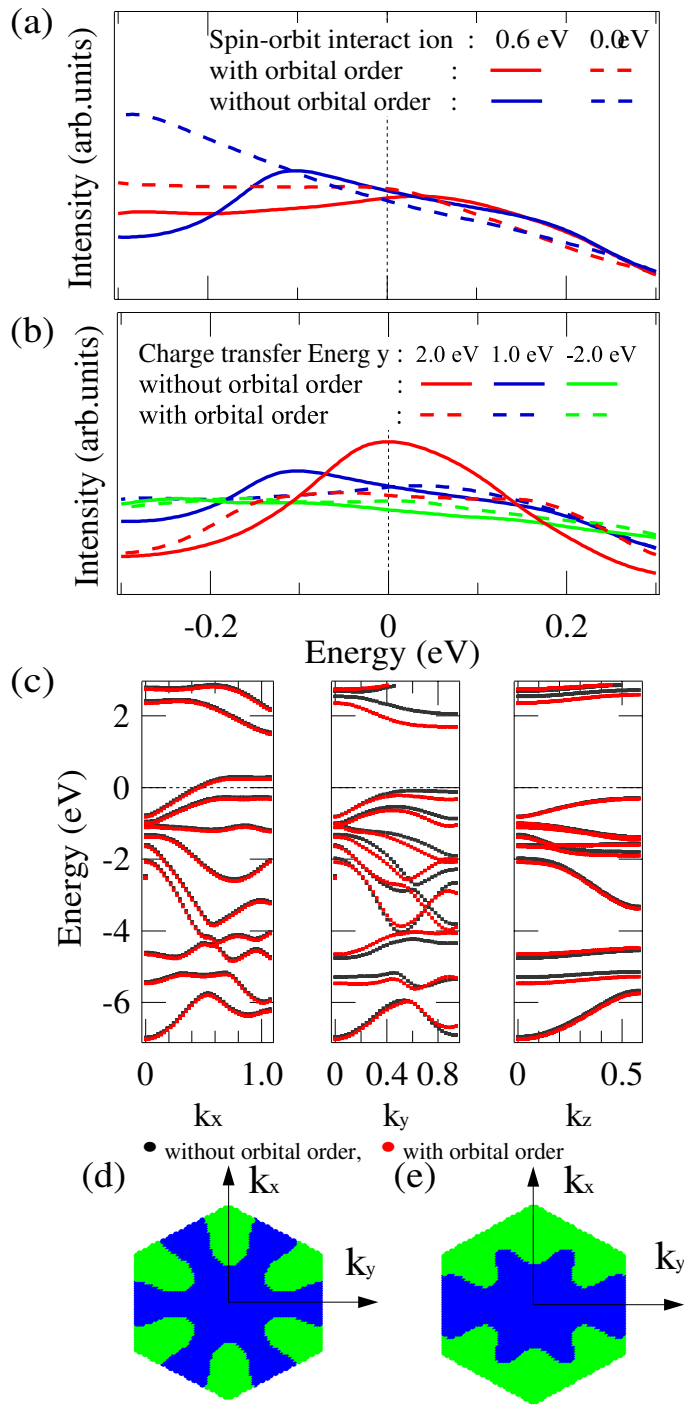}
\caption{(color online) (a) The effect of Ir 5$d$ spin-orbit interaction
on total density of states near the Fermi level calculated for IrTe$_2$ 
with and without orbital (or bond) order. $\Delta$ is set to 1.0 eV.
(b) Total density of states near the Fermi level calculated for IrTe$_2$
as functions of $\Delta$ with and without orbital (or bond) order.
(c) Band dispersions along the $k_x$, $k_y$, and $k_z$ directions.
$k_x$ and $k_y$ are parallel to the IrTe$_2$ plane,
and $k_z$ is perpendicular to it.
(d) Fermi surface without orbital (or bond) order. The Fermi surface
has the trigonal symmetry due to the orbital degeneracy of Ir 5$d$ 
yz, zx, and xy orbitals. The $k_x$ and $k_y$
directions are indicated in the figure. 
(e) Fermi surface with orbital (or bond) order.
The Fermi surface becomes quasi-one-dimensional due to 
a kind of band Jahn-Teller effect to remove the orbital 
degeneracy of Ir 5$d$ yz, zx, and xy orbitals. 
The $k_x$ and $k_y$ directions are indicated in the figure.
}
\end{figure}

With Pt substitution of $x \ge 0.032$, the orbital or bond order is suppressed 
and superconductivity appears.\cite{5} Actually, the spectral weight suppression 
around -0.1 eV of IrTe$_2$ almost disappears with the Pt doping. 
However, the dip structure around -0.1 eV slightly remains in the superconducting sample 
with $x=0.03$ and $0.04$ as shown in Figs. 3(b) and (c). Assuming that the dip structure 
around -0.1 eV is due to the orbital or bond order caused by the band Jahn-Teller effect, 
the band Jahn-Teller effect weakly affects the density of states of $x=0.03$ and $0.04$. 
Even in the superconducting sample with $x=0.04$, the dip structure still remains 
indicating that Ir$_{1-x}$Pt$_x$Te$_2$ has a kind of phase separation 
between the superconducting state and the orbital or bond order state.

As indicated by the Ir $4f$ XPS, the low temperature phase of IrTe$_2$ is accompanied by weak modulation 
of Ir $5d$ $t_{2g}$ electron density. The charge modulation or charge density wave can be induced 
by Fermi surface nesting due to the orbital or bond order (namely, due to the band Jahn-Teller effect). 
This situation is similar to the orbitally-induced Peierls effect proposed for CuIr$_2$S$_4$.\cite{14} 
In IrTe$_2$, when the orbital (or bond) order is established and the Ir $5d$ holes are accommodated 
in the Ir $5d$ yz orbitals, the Fermi surface is expected to become more one-dimensional to 
induce the charge density wave. The effect of orbital (or bond) order on the band dispersion 
and the Fermi surface geometry is demonstrated in Figs. 4(c), (d) and (e). 
Without the lattice distortion and the orbital (or bond) order, the Fermi surfaces 
of the Ir $5d$ bands are made up from the Ir $5d$ yz, zx, and xy orbitals and have 
the six-fold symmetry as expected from the trigonal structure. 
Under the compression along the b-axis and the orbital (or bond) order, 
the transfer integrals of the yz orbitals are enhanced and those of zx and xy orbitals 
are reduced. As a result, the Ir $5d$ band width along the $k_y$ direction is decreased due to the decrease of $zx$-$zx$ and $xy$-$xy$ transfer as shown in Fig. 4(c),
and the quasi-one-dimentional Fermi surface with yz character is obtained as shown in Fig. 4(e).
Such a quasi-one-dimensional Fermi surface is expected to have instability to charge 
or spin density wave. 
In contrast to CuIr$_2$S$_4$ \cite{10} and LiRh$_2$O$_4$ \cite{15} with full gap opening, 
the amplitude of the charge density wave is probably not enough to cause band gap opening in the case of IrTe$_2$. 

\section{Conclusion}

We have studied the electronic structure of a triangular lattice Ir$_{1-x}$Pt$_x$Te$_2$ 
superconductor. The combination between photoemission spectroscopy and model calculations show that 
the orbitally-induced Peierls effect on Ir $5d$ t$_{2g}$ bands plays important role in the charge 
and orbital instability in the IrTe$_2$. 
In the orbitally-induced Peierls mechanism, the Ir 5$d$ t$_{2g}$ orbital degeneracy 
is removed by a kind of band Jahn-Teller effect and the Fermi surface geometry is changed to enhance Fermi surface nesting.  
The structural change, the resistivity anomaly, and the spectral change 	
across the transition at $\sim$ 250 K in IrTe$_2$ is well explained by 
the orbitally-induced Peierls mechanism of the t$_{2g}$ electrons on the triangular lattice.
However, it is very difficult to explain the drastic suppression of magnetic susceptibility at the transition 
by the weak charge density wave and the imperfect gap opening. The effect of spin-orbit interaction, 
which is not included in the simple argument of the orbitally-induced Peierls mechanism, should be 
included to consider the remaining mystery. Also the suppression of magnetic susceptibility in IrTe$_2$ 
is very similar to that of LiVO$_2$, NaTiO$_2$, \cite{16} and LiVS$_2$ \cite{17} with $3d$ electrons
which are more localized than the $5d$ electrons in the Ir compounds. The effect of Mottness should be 
considered in future theoretical studies.

\section*{Acknowledgement}

The authors would like to thank valuable discussions with D. I. Khomskii and H. Takagi. 
The synchrotron radiation experiment was performed with the approval of HSRC (Proposal No.11-A-7).

\end{document}